\documentclass[12pt,letterpaper]{article}
\input{epsf}
\usepackage[pdftex]{graphicx,color}
\usepackage{hyperref}
\input{psfig}
\usepackage{amsmath,amssymb}
\usepackage{fancyhdr}
\usepackage{verbatim}
\newcommand\ltap{\
  \raise.3ex\hbox{$<$\kern-.75em\lower1ex\hbox{$\sim$}}\ }
\newcommand\gtap{\
  \raise.3ex\hbox{$>$\kern-.75em\lower1ex\hbox{$\sim$}}\ }

\newcommand\simge{\mathrel{%
   \rlap{\raise 0.511ex \hbox{$>$}}{\lower 0.511ex \hbox{$\sim$}}}}
\newcommand\simle{\mathrel{
   \rlap{\raise 0.511ex \hbox{$<$}}{\lower 0.511ex \hbox{$\sim$}}}}

\newcommand{\slashchar}[1]%
        {\kern .25em\raise.18ex\hbox{$/$}\kern-.70em #1}
\def\lsim{\mathrel{\raise.3ex\hbox{$<$\kern-.75em\lower1ex\hbox{$\sim$}}}}
\def\gsim{\mathrel{\raise.3ex\hbox{$>$\kern-.75em\lower1ex\hbox{$\sim$}}}}

\newcommand\CL{{\cal L}}
\newcommand\CM{{\cal M}}

\newcommand\CO{{\cal O}}

\newcommand\be{\begin{equation}}
\newcommand\ee{\end{equation}}
\newcommand\bea{\begin{eqnarray}}
\newcommand\eea{\end{eqnarray}}
\newcommand\ba{\begin{array}}
\newcommand\ea{\end{array}}
\newcommand\nn{\nonumber}

\newcommand{\thalf}{\textstyle{\frac{1}{2}}}

\newcommand\dagg{\dagger}

\newcommand\mev{{\rm MeV}}
\newcommand\gev{{\rm GeV}}
\newcommand\tev{{\rm TeV}}
\newcommand\MeV{{\rm MeV}}

\newcommand\etmiss{\slashchar{E}_T}
\newcommand\emiss{\slashchar{E}}

\newcommand\jet{{\rm jet}}

\begin{document}

\title{
\vskip -15mm
{\Large{\bf The 30 GeV Dimuon Excess at ALEPH}}\\
} \author{ {\large Kenneth Lane\thanks{lane@bu.edu} \,\,and Lukas
    Pritchett\thanks{lpritch@bu.edu}}\\
  {\large Department of Physics, Boston University}\\
  {\large 590 Commonwealth Avenue, Boston, Massachusetts 02215}\\
} \maketitle

\begin{abstract}

  A simple variation of a two-Higgs-doublet model is proposed to describe the
  $30\,\gev$ dimuon excess reported by Heister in his reanalysis of
  $Z \to \bar bb$ events in ALEPH data taken in 1992-95. The heavier CP-even
  Higgs $H$ is the $125\,\gev$ Higgs boson discovered at the LHC. The model
  admits two options for describing the dimuon excess: (1) The lighter
  CP-even Higgs $h$ and the CP-odd state $\eta_A$ are approximately
  degenerate and contribute to the $30\,\gev$ excess. (2) Only the $h$ is at
  $30\,\gev$ while the $\eta_A$ and $H$ are approximately degenerate at
  $125\,\gev$. The ALEPH data favor option~1. Testable predictions are
  presented for LHC as well as LEP experiments. A potential no-go theorem for
  models of this type is also discussed.

  \end{abstract}


\newpage

\section{Introduction}

\begin{figure}[!ht]
 \begin{center}
\includegraphics[width=2.60in, height=2.60in]{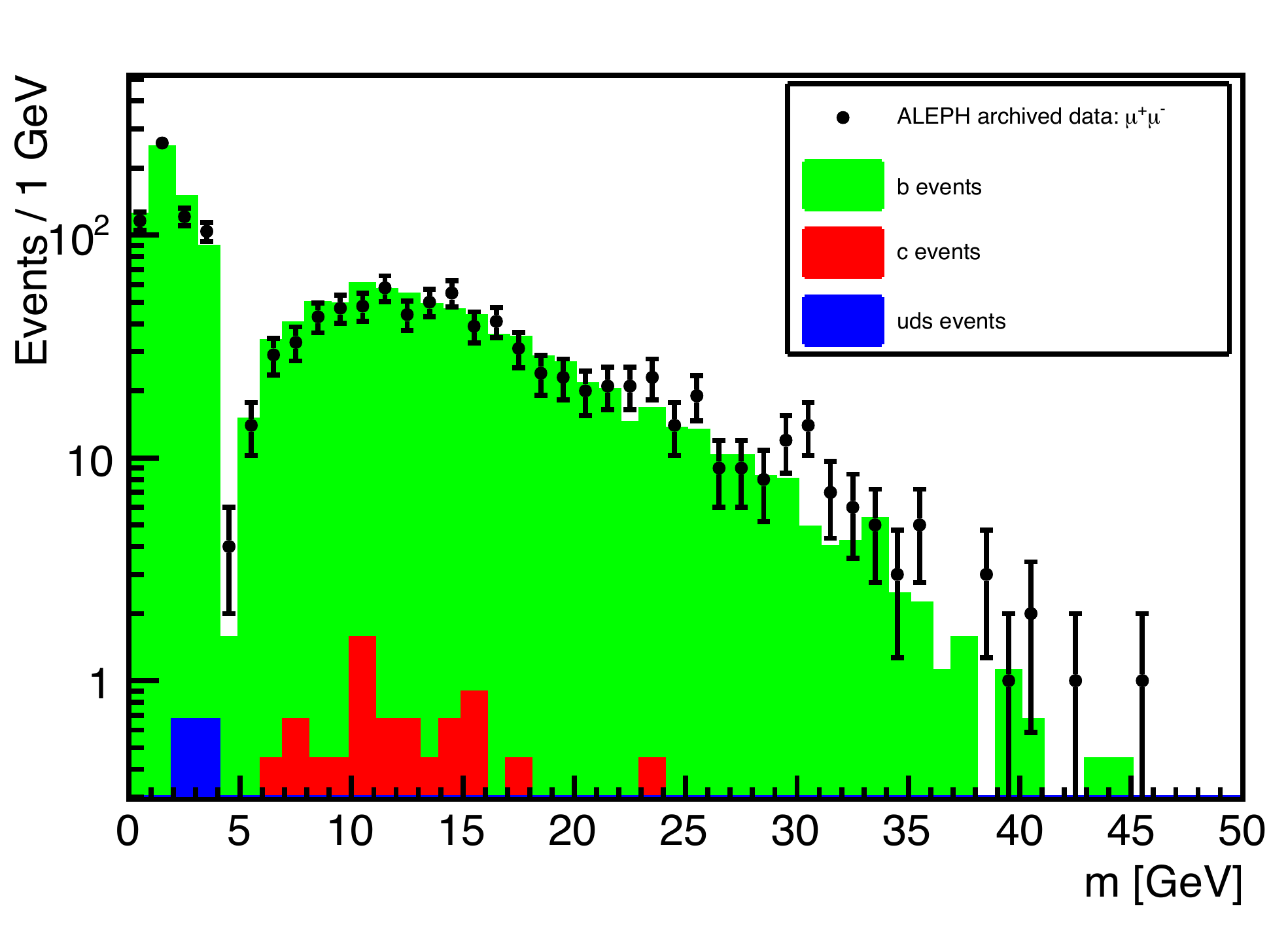}
\includegraphics[width=2.60in, height=2.60in]{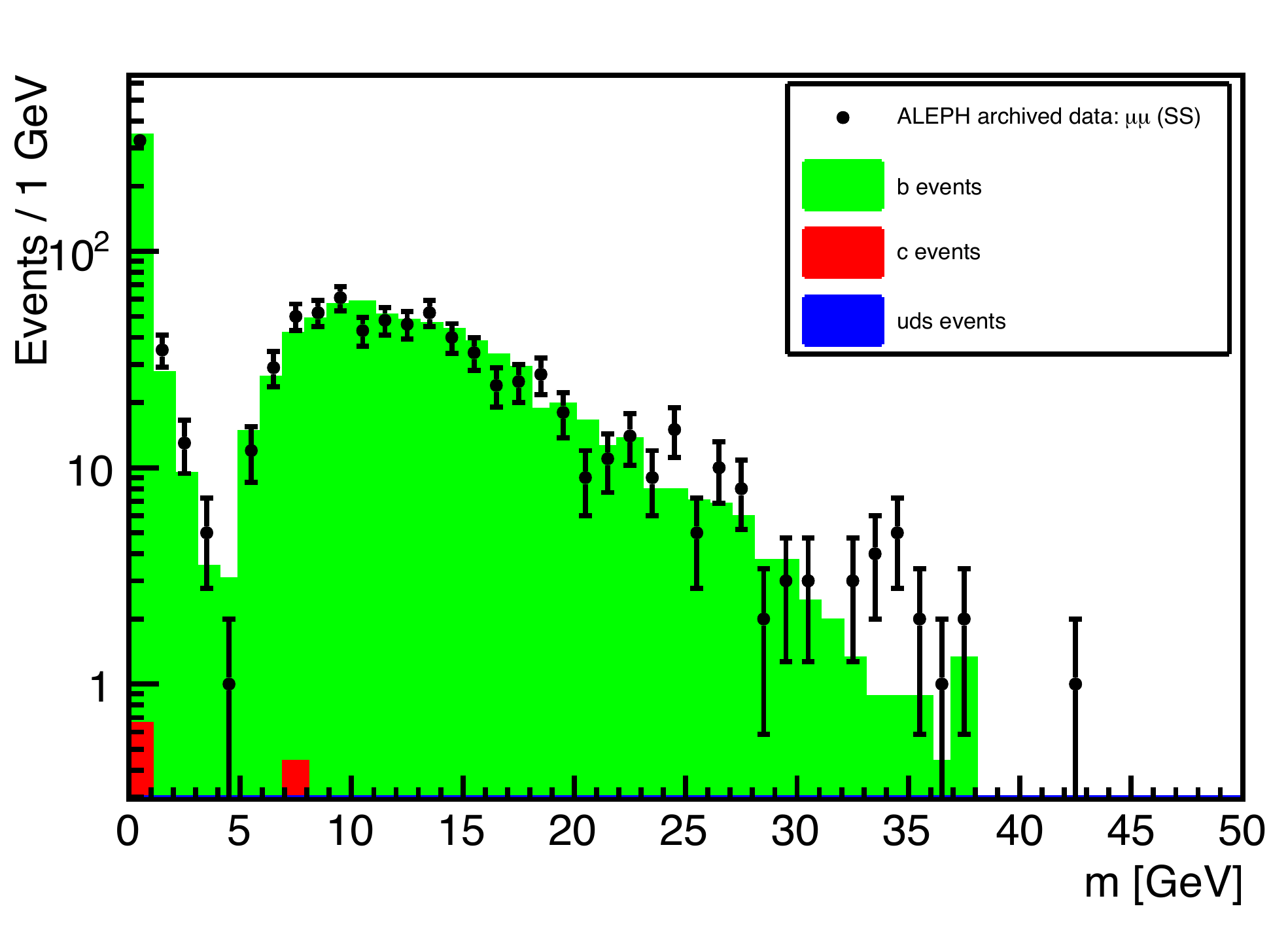}
\caption{The opposite-sign~(left) and same-sign~(right) dimuon mass spectra
  in $Z \to \bar bb \mu\mu$ data taken by the ALEPH Collaboration; from
  Ref.~\cite{Heister:2016stz}.}
  \label{fig:Fig1ab}
 \end{center}
 \end{figure}

In a recent paper, Heister analyzed archived data of the ALEPH experiment at
LEP and found apparent evidence for a narrow dimuon ($\mu^+ \mu^-$) resonance
at $30\,\gev$~\cite{Heister:2016stz}. The data, taken in 1992-95, involve
1.9~million hadronic decays of $Z$-bosons produced at rest in $e^+e^-$
annihilation. This excess appears in $Z \to \bar bb \mu^+\mu^-$ decays. The
opposite-sign dimuon spectrum data is shown in Fig.~1a along with the
expected background. The same-sign dimuon spectrum in Fig.~1b has no
significant excesses. The data have the following characteristics:
\begin{itemize}

\item[1.)] Two benchmark methods were used to estimate the significance of
  the excess. One gave a local significance of about $2.6\,\sigma$, the other
  $5.4\,\sigma$. The second method requires using the look-elsewhere effect;
  it reduces its significance by~1.4--$1.6\,\sigma$. See
  Ref.~\cite{Heister:2016stz} for details.

\item[2.)] There is an excess of $32\pm 11$ events in the resonant peak of
  Fig.~2 corresponding to a mass of $30.40\,\gev$ with a Breit-Wigner width
  of $1.78\,\gev$ (Gaussian width of $0.74\,\gev$), consistent with the
  expected ALEPH dimuon mass reconstruction performance at $30\,\gev$. Using
  the $b$-tag and muon-ID efficiencies quoted in Ref.~\cite{Heister:2016stz},
  this yields the branching ratio
\be\label{eq:BZbbmm}
B(Z \to \bar b b\, X (\to \mu^+ \mu^-)) = (2.77 \pm 0.95)\times 10^{-4}.
\ee
It should be understood that, if the dimuon excess is due to the decay of a
new particle $X$, it is not known whether it is emitted from the $Z$, as in
$Z \to Z^* X$ with $Z^* \to \bar b b$ and $X \to \mu^+\mu^-$, or from one of
the $b$-quarks, as in $Z \to \bar bb \to \bar bb + X$, or from two new
particles, $Z \to XY$, with $X \to \mu^+ \mu^-$ and $Y \to \bar bb$.

\begin{figure}[!t]
 \begin{center}
\includegraphics[width=3.15in, height=3.15in]{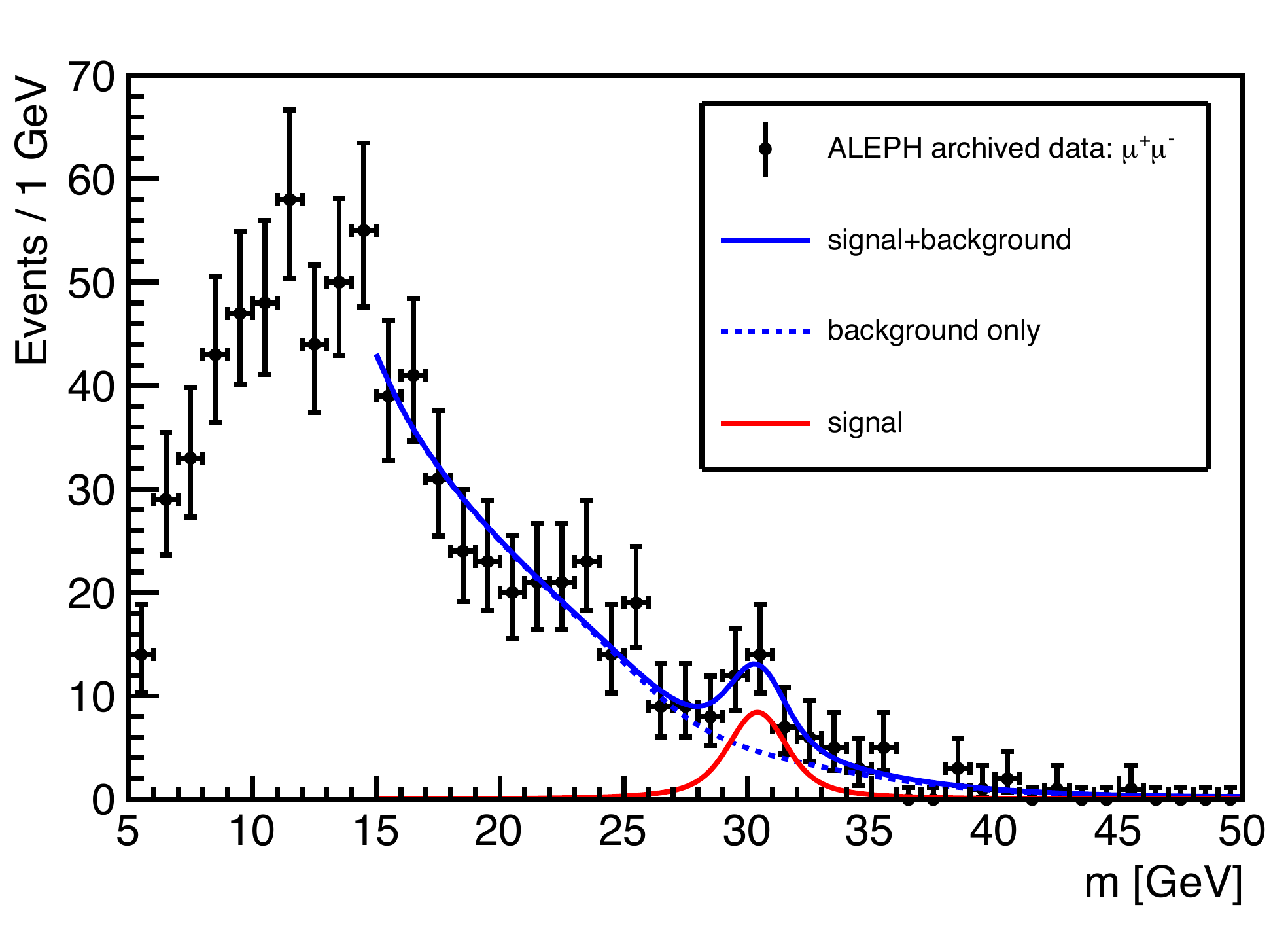}
\caption{ALEPH $Z \to \bar bb \mu^+\mu^-$ data with signal$+$background model
  used to extract the 30~GeV signal parameters in
  Ref.~\cite{Heister:2016stz}.}
  \label{fig:Fig2}
 \end{center}
 \end{figure}

\item[3.)] The decay angle ($\cos\theta^*$) distribution for muons in the
  dimuon rest frame, where $\theta^*$ is the angle between the dimuon boost
  axis and the~$\mu^-$, is shown in Fig.~3a for the signal region, a mass
  range of $2\,\sigma$ around the fitted mean mass value,
  $M_{\mu^+\mu^-} = (30.40 \pm 3.85)\,\gev$. There is a clear preference for
  forward-backward production, i.e., with each muon close to a
  $b$-jet. Presumably, most of these events are semileptonic
  $b$-decays. There is also a smaller, approximately isotropic component for
  $|\cos\theta^*| < 0.8$. This may be indicative of a different -- scalar --
  production mechanism in the signal region. However, Fig.~3b shows the
  angular distribution of events in sidebands, with
  $M_{\mu^+\mu^-} = 15$--$50\,\gev$ but excluding the signal-region events of
  Fig.~3a. It does not appear substantially different from Fig.~3a (though
  the ratio of events at $|\cos\theta^*| > 0.8$ to those in between is
  greater than it is in Fig.~3a).

\begin{figure}[!t]
 \begin{center}
\includegraphics[width=2.60in, height=2.60in]{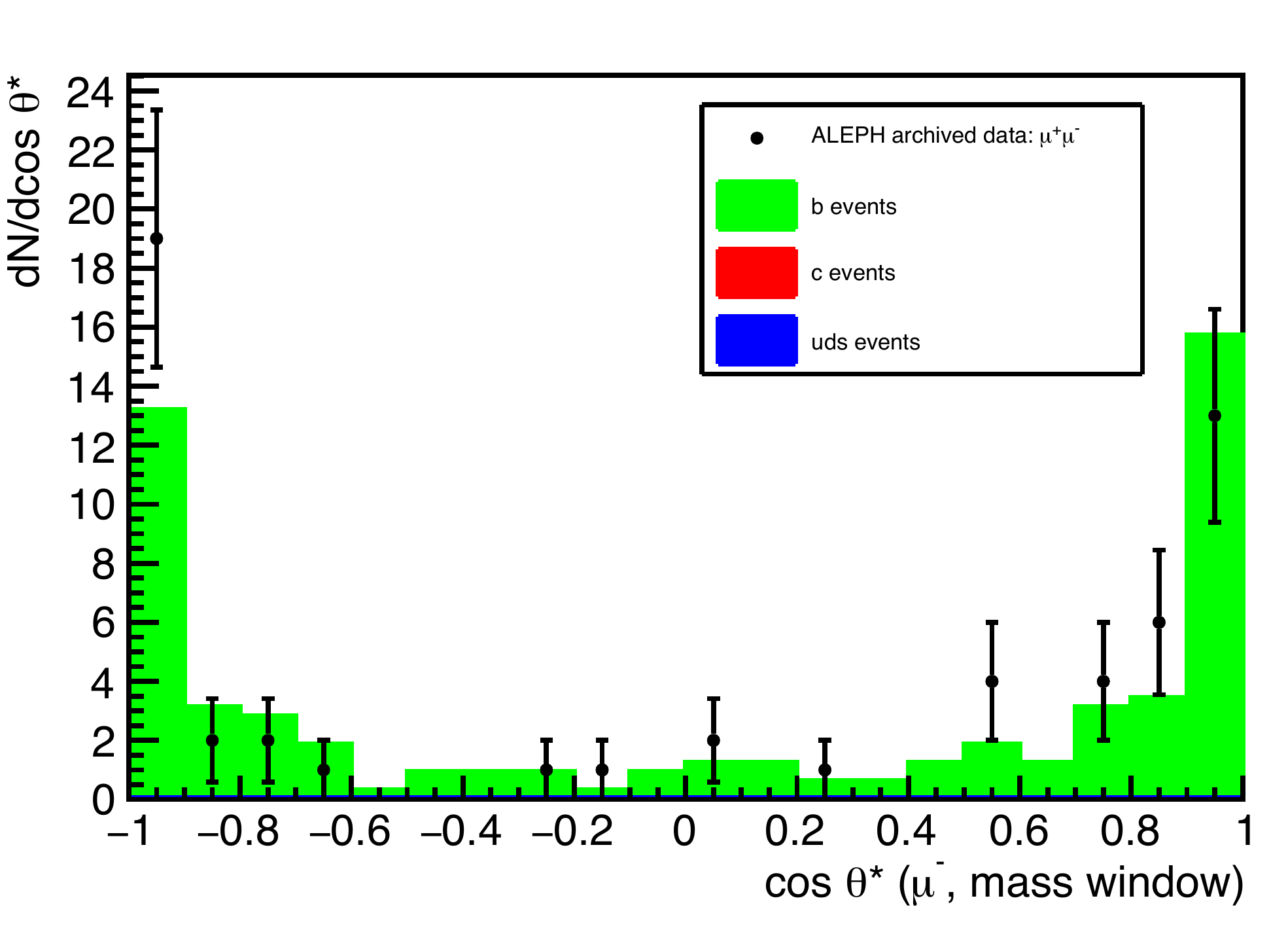}
\includegraphics[width=2.60in, height=2.60in]{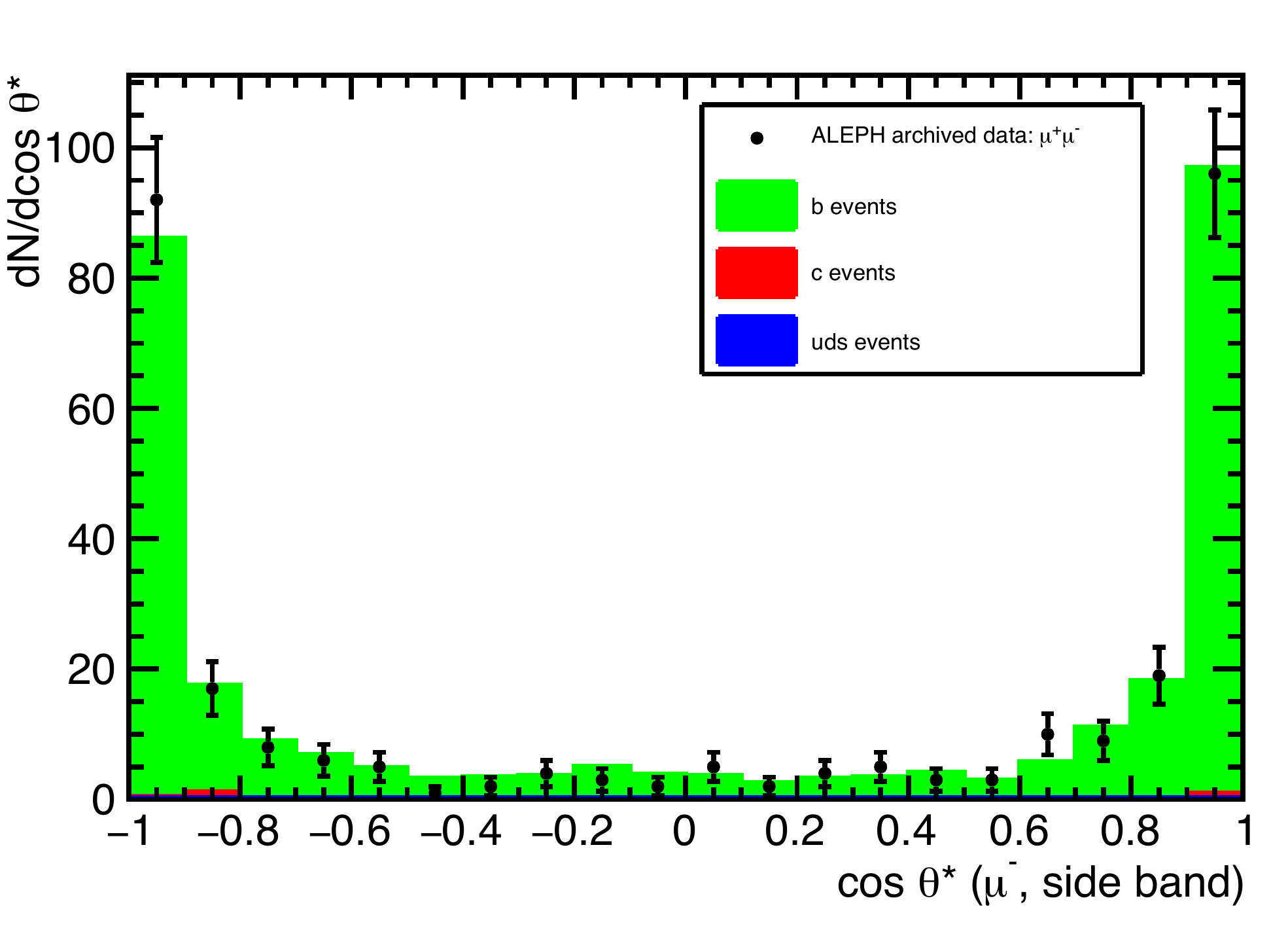}
\caption{The decay angle ($\cos\theta^*$) distribution for muons (left) in the
  signal region, $M_{\mu^+\mu^-} = (30.40 \pm 3.85)\,\gev$ and (right) in the
  sidebands $15 < M_{\mu^+\mu^-} < 50\,\gev$, excluding the signal region;
  from Ref.~\cite{Heister:2016stz}.}
  \label{fig:Fig3ab}
 \end{center}
 \end{figure}

\item[4.)] As noted above, there is no significant excess near
  $M_{\mu\mu} = 30\,\gev$ excess in the same-sign data,
  $Z \to \bar bb \mu^\pm \mu^\pm$. Nor is there an excess in the
  opposite-sign electron-muon data, $Z \to \bar bb e^\pm \mu^\mp$.

\item[5.)] There is a small excess of $8.0 \pm 4.5$ events near $M_{e^+ e^-}
  = 30\,\gev$ in the $Z \to \bar bb e^+ e^-$ data.

\item[6.)] There is no evidence for the 30~GeV dimuon excess in events for
  which the $b$-tag has been inverted; see Fig.~4 from
  Ref.~\cite{Heister:2016stz}. Comparison of Fig.~4 with Fig.~1a shows that
  most of the events near $M_{\mu^+\mu^-} = 30\,\gev$ are still $\bar bb$, so
  it is not clear how dispositive this is of the excess being produced only
  in association with $\bar bb$.

\item[7.)] Ref.~\cite{Heister:2016stz} states that, for $M_{\mu^+\mu^-}$ in
  the vicinity of $30\,\gev$, the minimum angle between one of the two muons
  and the leading jet was always found to be less than
  $15^\circ$.\footnote{Ref.~\cite{Heister:2016stz} also observed a tendency
    for at least one of the leading jets to be broadened when the dimuon mass
    is high. This may make it difficult for to define the $b$-jet axis
    precisely in such events.}

\end{itemize}

\begin{figure}[!t]
 \begin{center}
\includegraphics[width=3.15in, height=3.15in]{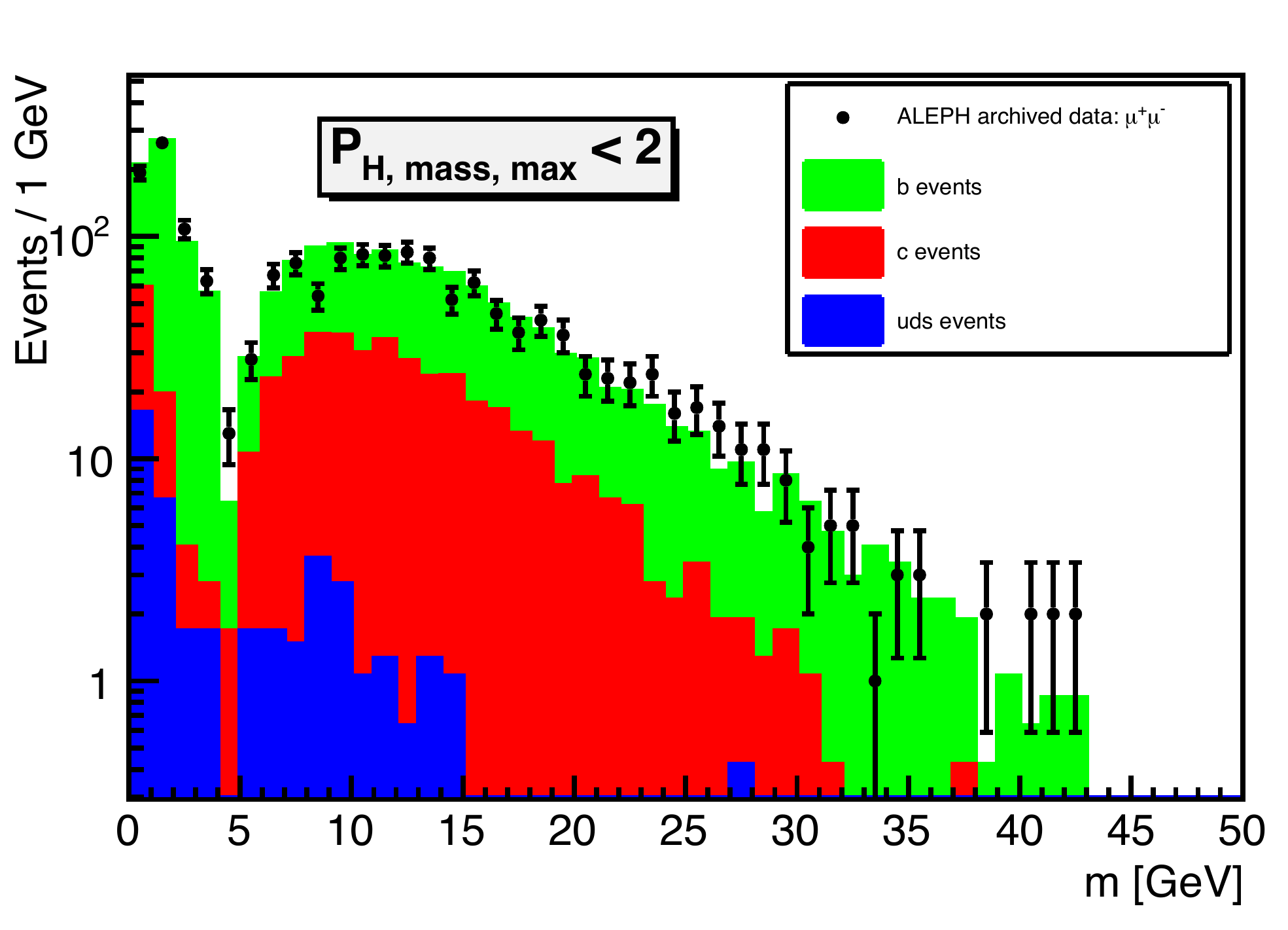}
\caption{The opposite-sign dimuon mass spectrum in $Z \to {\rm hadrons}+\mu^+
  \mu^-$ events in which the $b$-tag has been inverted, indicating no evidence
  for an excess near $30\,\gev$; from Ref.~\cite{Heister:2016stz}.}
  \label{fig:Fig4}
 \end{center}
 \end{figure}

 The obvious and simplest explanation of these features of the ALEPH data is
 that the 30~GeV excess is just a statistical fluctuation in semileptonic
 $Z \to \bar bb$ decays. On the other hand, it is possible to construct a
 rather minimal model that accounts for the ALEPH data and makes several
 testable predictions. It is a two-Higgs doublet model (2HDM) in which the
 heavier CP-even Higgs boson $H$ is the $125\,\gev$ Higgs boson discovered in
 2012 at the LHC~\cite{Aad:2012tfa, Chatrchyan:2012ufa}. The two other
 neutral Higgs bosons are a CP-even one~$h$ and a CP-odd one~$\eta_A$. The
 additional neutral and charged Higgs bosons couple mainly to the muon
 doublet and secondarily, but more weakly, to $b$-quarks. We shall choose
 parameters so that $M_h = 30\,\gev$. There are then two ``natural'' options
 for the $\eta_A$; either (1)~$M_{\eta_A} \cong M_h = 30\,\gev$ or
 (2)~$M_{\eta_A} \cong M_H = 125\,\gev$. In option~1, $Z \to h \eta_A$ with
 $h \to \mu^+ \mu^-,\,\, \eta_A \to \bar bb$ and vice-versa. There are also
 two ``Higgsstrahlung'' processes: $Z \to Z^* h$ with $Z^* \to \bar bb$
 and~$h \to \mu^+\mu^-$; and $Z \to \bar bb$ with~$b$ or~$\bar b$
 radiating~$h$ or~$\eta_A$ which then decays to $\mu^+\mu^-$. In option~2,
 there are only the Higgsstrahlung processes involving $h$-radiation. The
 branching ratio~(\ref{eq:BZbbmm}) is easily fit by the first option, but not
 the second. If the charged Higgs bosons in this model, $h^\pm$, are heavier
 than $M_H/2$, they may have evaded previous searches because they decay
 mainly to $\mu^\pm \nu_\mu$ and rarely to $\tau^\pm \nu_\tau$, $c\bar b$,
 and $c\bar s$; see, e.g., Refs.~\cite{Olive:2016xmw,Khachatryan:2015qxa,
   Aad:2014kga} for $h^\pm$~and other searches at LEP and the LHC.

 The rest of this paper is organized as follows: In Sec.~2 we describe our
 2HDM model: its assumptions and their rationale; its potential, extremal
 conditions and mass matrices; the Higgs couplings to leptons, quarks,
 electroweak gauge bosons and to each other. In Sec.~3 we present the two
 options for describing the 30~GeV dimuon excess. There we see that only
 option~1 can explain Eq.~(\ref{eq:BZbbmm}) and we present numerical values
 for the model's parameters and the corresponding signal branching ratio of
 the $Z$. Sec.~4 catalogs predictions of our model. Some of these may be
 useful for looking for the dimuon in LHC experiments. Finally, in Sec.~5, we
 present what appears to be a fatal flaw of the model, and a potential no-go
 theorem for any Higgs-based (and other scalar-based) model of the 30~GeV
 dimuon. However, if the excess seen in ALEPH is confirmed in other LEP and
 LHC experiments, it will be difficult to dismiss the dimuon as a background
 fluctuation and this fly in the ointment will stand as a significant
 challenge to model-builders.

\section{The 2HDM model}

The model uses the two Higgs doublets (see Ref.~\cite{Branco:2011iw} for a
review),
\be\label{eq:higgs}
\phi_i = \frac{1}{\sqrt{2}}\left(\begin{array}{c} \sqrt{2}\phi^+_i \\
\phi_{i0} - i \phi_{i3}\end{array}\right),
\ee
where $\phi^\pm_i = \textstyle{1/\sqrt{2}}(\phi_{i1} \mp i \phi_{i2})$ for
$i=1,2$. Both doublets have weak hypercharge $\thalf$. To account for the
appearance of a dimuon excess only in association with the $\bar bb$ decays
of $Z$-bosons, we assume a $U(1)_\phi$ symmetry with $\phi$-hypercharge
$Y_\phi$ assignments for the Higgs doublets, left handed fermion doublets and
right-handed fermion singlets as follows:\footnote{This is a simple version
  of the Branco-Grimus-Lavoura models of Ref.~\cite{Branco:1996bq} which has
  no Higgs-induced flavor-changing neutral current interactions; also see
  Refs.~\cite{Botella:2014ska, Bhattacharyya:2014nja}.}
\bea\label{eq:reflect}
&& Y_\phi(\phi_1) = 0, \,\, Y_\phi(\phi_2) = 1; \nn\\
&& Y_\phi(q_{Lk})  = Y_\phi\left(\begin{array}{c}u_{Lk}\\ d_{Lk}\end{array}\right) =
Y_\phi(u_{Rk}) = Y_\phi(d_{Rk}) =0 ; \quad (k=1,2,3) \nn\\
&& Y_\phi(L_{Lk}) = Y_\phi\left(\begin{array}{c}\nu_{Lk}\\
                                  \ell_{Lk}\end{array}\right) = \thalf,
\quad Y_\phi(\ell_{Rk}) = -\thalf; \,\,\,\,\qquad (k=1,2) \nn\\
&& Y_\phi(L_{L3}) = Y_\phi(\ell_{R3}) = 0 .
\eea
This symmetry is softly broken by the dimension-two $\phi_1^\dagg \phi_2$ term
in the potential\footnote{The quartic $\lambda$-couplings in
  Eq.~(\ref{eq:Vphi}) are half the corresponding ones in
  Ref.~\cite{Branco:2011iw}. The $\lambda_5$ term in that reference is
  forbidden here by the (softly-broken) $U(1)_\phi$ symmetry.}
\bea\label{eq:Vphi}
V(\phi_1,\phi_2) &=&  -\mu_1^2\, \phi_1^\dagg \phi_1  -\mu_2^2\, \phi_2^\dagg
\phi_2  -\mu_3^2\, (\phi_1^\dagg \phi_2 + \phi_2^\dagg \phi_1) + \lambda_1
(\phi_1^\dagg \phi_1)^2 \nn\\
&& \,\, + \lambda_2 (\phi_2^\dagg
\phi_2)^2  + 2\lambda_3 (\phi_1^\dagg \phi_1)(\phi_2^\dagg \phi_2)
 + 2\lambda_4 (\phi_1^\dagg \phi_2)(\phi_2^\dagg \phi_1).
\eea
Here, $\mu^2_{1,2,3} > 0$, all $\lambda$'s are real, $\lambda_{1,2} > 0$ for
vacuum stability, and we will want to assume that $\lambda_4 < 0$. For a
range of these parameters, then, these fields have the real vacuum
expectation values (vevs)
\be\label{eq:vevs}
\langle\phi_i\rangle_0 =
\frac{1}{\sqrt{2}}\left(\begin{array}{c} 0 \\ v_i\end{array}\right),
\ee
and they satisfy the extremal conditions
\bea\label{extrema}
&& -\mu_1^2 -\mu_3^2\,v_2/v_1 + \lambda_1 v_1^2 +(\lambda_3 + \lambda_4)v_2^2
= 0,\\
&& -\mu_2^2 -\mu_3^2\,v_1/v_2 + \lambda_2 v_2^2 +(\lambda_3 + \lambda_4)v_1^2
= 0.
\eea
The square of the electroweak vev is
$v^2 = v_1^2 + v_2^2 = (246\,\gev)^2$. The mass matrices, mass eigenstate
fields and eigenvalues of the CP-even Higgs bosons are (after shifting them
by their respective vevs):
\bea\label{eq:MHh}
M^2(\phi_{10},\phi_{20}) &=& \left(\begin{array}{cc}
\mu_3^2\,v_2/v_1 + 2\lambda_1 v_1^2 & -\mu_3^2 +2(\lambda_3 + \lambda_4)v_1v_2
                                   \\
-\mu_3^2 +2(\lambda_3 + \lambda_4)v_1v_2 & \mu_3^2\,v_1/v_2 + 2\lambda_2 v_2^2
\end{array}\right),
\\ \nn\\
H &=& \phi_{10}\cos\alpha + \phi_{20}\sin\alpha,\quad
h = -\phi_{10}\sin\alpha + \phi_{20}\cos\alpha,\nn\\
{\rm where\,\,}
\tan 2\alpha &=& \frac{2\left[2(\lambda_3 + \lambda_4)v_1 v_2 - \mu_3^2\right]}
                    {\mu_3^2(v_2/v_1 - v_1/v_2) + 2(\lambda_1 v_1^2 -
                      \lambda_2 v_2^2)}; \\
M^2(H,h) &=& \frac{1}{2v_1 v_2}\biggl\{\mu_3^2 v^2 + 2(\lambda_1 v_1^2 +
  \lambda_2 v_2^2)v_1 v_2 \\
&\quad& \pm\biggl[\bigl[\mu_3^2 v^2 + 2(\lambda_1 v_1^2 + \lambda_2 v_2^2)v_1
v_2)\bigr]^2
             -8\bigl[\mu_3^2(\lambda_1 v_1^4 + \lambda_2 v_2^4)v_1v_2 \nn \\
&\qquad& \,\,  + 2\lambda_1\lambda_2 v_1^4 v_2^4
               +2\mu_3^2(\lambda_3 + \lambda_4)v_1^3 v_2^3 -2(\lambda_3 +
               \lambda_4)^2 v_1^4 v_2^4\bigr]\biggr]^{\thalf}\biggr\}.\nn
\eea
For the CP-odd Higgs bosons, they are:
\bea\label{eq:Mpieta}
M^2(\phi_{13},\phi_{23}) &=& \left(\begin{array}{cc}
\mu_3^2v_2/v_1 & -\mu_3^2 \\
-\mu_3^2 & \mu_3^2v_1/v_2
\end{array}\right),
\\ \nn\\
\pi_A &=& \phi_{13}\cos\beta + \phi_{23}\sin\beta,\quad
\eta_A = \phi_{13}\sin\beta - \phi_{23}\cos\beta,\nn\\
{\rm where\,\,}
\tan\beta &=& \frac{v_2}{v_1}; \\
M^2_{\pi_A} &=& 0,\quad M^2_{\eta_A} = \mu_3^2\frac{v^2}{v_1 v_2}.
\eea
The $\eta_A$ is a pseudo-Goldstone boson of the spontaneously broken
$U(1)_\phi$ symmetry which is also softly broken by the $\mu_3^2$-term in
the Higgs potential. For the charged Higgs bosons:
\bea\label{eq:Mpipm}
M^2(\phi^\pm_{1},\phi^\pm_{2}) &=& \left(\begin{array}{cc}
\mu_3^2v_2/v_1- \lambda_4 v_2^2 & -\mu_3^2 + \lambda_4 v_1 v_2\\
-\mu_3^2 + \lambda_4 v_1 v_2 & \mu_3^2v_1/v_2 - \lambda_4 v_1^2
\end{array}\right);
\\ \nn\\
\pi^\pm &=& \phi^\pm_{1}\cos\beta + \phi^\pm_{2}\sin\beta,\quad
h^\pm = \phi^\pm_{1}\sin\beta - \phi^\pm_{2}\cos\beta, \\
M^2_{\pi^\pm} &=& 0, \quad M^2_{h^\pm} = \biggl(\frac{\mu_3^2}{v_1 v_2}
-\lambda_4\biggr)v^2.
\eea

To be consistent with the ALEPH data, we assume that the scalar
doublet~$\phi_1$ couples to all fermions except the muon and electron, while
$\phi_2$ couples only to the~$\mu$ and~$e$ doublets.\footnote{Alternatively,
  we could just as well couple the electron to $\phi_1$.} As noted above,
this is implemented by the (softly-broken) $U(1)_\phi$ symmetry on Higgs and
fermion fields. Without loss of generality, the Yukawa terms for the leptons
may then be written in terms of mass-eigenstate lepton fields as
%
%
\bea\label{eq:lepty}
\CL_{Y\ell} &=& -\sum_{\ell_k=e,\mu} \frac{m_{\ell_k}}{v\sin\beta}
                  \,\bar \ell_k \left[v\sin\beta + H\sin\alpha + h\cos\alpha
                    +i\eta_A\gamma_5\cos\beta\right]\, \ell_k
                  \nn\\
&& -\frac{m_\tau}{v\cos\beta} \,\bar\tau\left[v\cos\beta + H\cos\alpha -
  h\sin\alpha - i\eta_A\gamma_5\sin\beta \right]\,\tau \\
&&  + h^+\left[\sum_{k=e,\mu}
    \frac{\sqrt{2}m_{\ell_k}\cot\beta}{v}\,\bar\nu_{k_L} \ell_{kR}
  - \frac{\sqrt{2}m_\tau    \tan\beta}{v}\,\bar\nu_{\tau L}\, \tau_R\right]
    + {\rm h.c.}  \nn
\eea
{\em These interactions induce no detectable charged-lepton flavor
  violation.}\footnote{The $h^\pm$-contribution to the rate for
  $b \to s\gamma$ is suppressed by $\tan^4\beta$.} The Yukawa interactions of
the quarks are
\bea\label{eq:quarky}
\CL_{Yq} &=& -\sum_{d_k=d,s,b} \frac{m_{d_k}}{v\cos\beta}
                  \,\bar d_k \left(v\cos\beta + H\cos\alpha -h\sin\alpha
                    -i\eta_A\gamma_5\sin\beta \right) d_k
                  \nn\\
&& -\sum_{u_k=u,c,t} \frac{m_{u_k}}{v\cos\beta}\,
               \bar u_k \left(v\cos\beta + H\cos\alpha -h\sin\alpha
                 +i\eta_A\gamma_5\sin\beta \right) u_k  \\
&& -\frac{\sqrt{2} \tan\beta}{v} \sum_{k,l=1}^3
\left[\bar u_{kL} (V\CM_d)_{kl}\, h^+ \,d_{lR} -
      \bar d_{kL} (V^\dagg\CM_u)_{kl} \, h^- \, u_{lR}\right] + {\rm h.c.}\nn
\eea
Here, $\CM_{u,d}$ are the diagonal up and down-quark matrices and $V$ is the
Cabibbo-Kobayashi-Maskawa (CKM) matrix. For small~$\alpha$ and~$\beta$, $h$
and $\eta_A$ decay mainly to $\mu^+ \mu^-$ and, at most at the percent level,
to $\bar bb$. The $h^\pm$ decay almost entirely to $\mu^\pm \nu_\mu$. Because
of this, the limits on charged Higgses from $Z$~and $t$-decay appear to be
inapplicable because they assume
$h^\pm \to \tau^\pm \nu_\tau,\, c \bar b,\, c\bar s$~\cite{Olive:2016xmw},
modes with very small branching ratios in our model. In Sec.~3, we shall find
it prudent to assume $M_{h^\pm} > M_H/2$, hence $\lambda_4 < 0$.

The most important couplings of the Higgses to electroweak bosons are (in
unitary gauge):
\bea\label{eq:phiEW}
\CL_{EW} &=& \frac{e}{\sin 2\theta_W}\left[\left(h\cos(\beta-\alpha) -
  H\sin(\beta-\alpha)\right)\overleftrightarrow{\partial_\mu}\,\eta_A\right]
Z^\mu \nn\\ 
&+& \frac{e}{2\sin\theta_W}\left[\left(\eta_A \pm i h\cos(\beta-\alpha) \mp
  i H\sin(\beta-\alpha)\right)\overleftrightarrow{\partial_\mu}\,h^\pm \right]
W^{\mp\,\mu} \\
&+& \left[\frac{2e^2}{\sin^2 2\theta_W} Z^\mu Z_\mu +
    \frac{e^2}{\sin^2\theta_W} W^{+\,\mu} W_\mu^-\right]
  \left[H\cos(\beta-\alpha) + h\sin(\beta-\alpha)\right]. \nn
\eea
For small $\alpha$ and $\beta$, the couplings of~$H$ are close to the
Standard Model (SM) in all cases. Note the strong $Z \to h\eta_A$ coupling.

Finally, for light $h$, $\eta_A$ and $h^\pm$, there is the possibility of
$H$-decay to pairs of them. The relevant Lagrangian for this is:
\bea\label{eq:Hhh}
\CL_{H\phi\phi} &=& vHh^2\bigl[3(\lambda_1 c_\beta c_\alpha s^2_\alpha +
                            \lambda_2 s_\beta s_\alpha c^2_\alpha)\nn\\ 
&+& (\lambda_3 + \lambda_4)(c_\beta c_\alpha(1-3s^2_\alpha) +
                        s_\beta s_\alpha(1-3c^2_\alpha))\bigr] \nn\\
  &+& vH\eta_A^2 \bigl[\lambda_1 c_\beta c_\alpha s^2_\beta + \lambda_2
  s_\beta s_\alpha c^2_\beta 
  + (\lambda_3 + \lambda_4)(c^3_\beta c_\alpha + s^3_\beta
  s_\alpha) \bigr] \nn\\
  &+& 2vH h^+ h^-
\bigl[\lambda_1 c_\beta c_\alpha s^2_\beta + \lambda_2 s_\beta s_\alpha
  c^2_\beta  
  + \lambda_3 (c^3_\beta c_\alpha + s^3_\beta s_\alpha) \nn\\
  &-& \lambda_4 c_\beta s_\beta \sin(\beta+\alpha)\bigr].
\eea
where $c_\beta = \cos\beta$, etc.

\section{Options for the 30 GeV Dimuon Excess}

We identify $H$ as the 125~GeV Higgs boson and $h$ and possibly $\eta_A$ as
the 30~GeV excess in Ref.~\cite{Heister:2016stz}. In order that this be
consistent with LHC data on~$H$, particularly the Higgs signal
strengths~\cite{Olive:2016xmw}, we require rather weak coupling between
$\phi_1$ and $\phi_2$. This means small~$\alpha$ for $\phi_{10}$--$\phi_{20}$
mixing and small $\beta$ for mixing of the CP-odd scalars and of the charged
scalars, i.e.,
\be\label{eq:assume}
v^2 \cong v_1^2 \gg v_2^2.
\ee
Then we can make the further reasonable assumption that
$(\mu_3^2 - 2\lambda_1 v_1v_2)^2 \gg 8(\lambda_1 \lambda_2 -(\lambda_3 +
\lambda_4)^2) v_2^4$, and we obtain
\be\label{eq:M2Hh}
M^2(H,h) \cong \left\{\begin{array}{c}
{\rm max}(2\lambda_1 v_1^2,\,\mu_3^2 v^2/v_1 v_2)\\
{\rm min}(2\lambda_1 v_1^2,\,\mu_3^2 v^2/v_1 v_2)\\
             \end{array}\right..
\ee
Thus, there are two options for the extra Higgs bosons'
masses:\footnote{Option 1 is the same as considered in
  Ref.~\cite{Bernon:2015wef}, except that we forbid the $\lambda_{5,6}$
  quartic couplings of the 2HDM.}
\bea\label{eq:options}
(1) \quad M^2_H &\cong& 2\lambda_1 v^2\,\,{\rm and}\,\, M^2_h \cong
M^2_{\eta_A} \cong \mu_3^2 v^2/v_1 v_2 ;\\ \nn\\
(2) \quad M^2_H &\cong& M^2_{\eta_A} \cong \mu_3^2 v^2/v_1 v_2\,\,{\rm
  and}\,\, M_h^2 \cong 2\lambda_1 v^2 .
\eea
The solution $M_H^2 = 2\lambda_1 v_1^2$ is the SM formula for the Higgs
boson's mass. If this option is preferred by the ALEPH data, then $h$ and
$\eta_A$ are nearly degenerate. In either case, the charged Higgs boson mass,
$M^2_{h^\pm} = \mu_3^2 v^2/v_1 v_2 - \lambda_4 v^2 = M_{\eta_A}^2 - \lambda_4
v^2$, depends on the sign and magnitude of $\lambda_4$. 

For small $\beta$ there is not much leeway in the masses of these two
options. In option~1, making $M_{\eta_A} < M_h$ quickly leads to $\lambda_2$
an order of magnitude larger than $\lambda_1$ and potentially to trouble with
$H$ decays to the light scalars (see below). Anyway, there is little
motivation for $M_{\eta_A} < M_h$. Making $M_{\eta_A} > M_h$ even more
quickly leads to $\lambda_2 < 0$ and an unstable Higgs potential.

Another feature of option~1 is that $H$ can decay to $hh$, $\eta_A \eta_A$
and $h^+ h^-$. A glance at Eq.~(\ref{eq:Hhh}) shows that these decays are
strongly dominated by the $\CO(\cos\beta \cos\alpha)$ terms in the
$\lambda_3$ and $\lambda_4$ interactions; for moderate values of $\lambda_1$
and $\lambda_2$, their interactions contribute negligibly to the Higgs width. For
$|\lambda_3|,\,|\lambda_4| =$ few $\times 10^{-2}$, these processes
contribute several 10's of~MeV to the Higgs width, an order of magnitude more
than its SM width of $4.07\,\mev$. We choose $\lambda_3 + \lambda_4$ so that
the $hh$ and $\eta_A \eta_A$ contributions to the Higgs width are each
$\simle \thalf\,\MeV$. This implies
$|\lambda_3 + \lambda_4| \simle 5.44\times 10^{-3}$ and, in turn, small
values and range for $\alpha$.\footnote{This and other such constraints on
  2HDM quartic couplings were discussed in
  Ref.~\cite{Bernon:2014nxa}.}$^,$\footnote{This constraint is consistent
  with limits on $H$-decays to light bosons in
  Ref.~\cite{Khachatryan:2017mnf}.} For example,
\be
\label{eq:alphrange} -4.30\times 10^{-3} \simle \alpha \simle -0.671\times
10^{-3} \quad {\rm for}\,\, v_2 = 10\,\gev.
\ee
In addition, we shall take $\lambda_4 < 0$ so that
$M_{h^\pm} = 65\,\gev > M_H/2$.\footnote{It is possible that the main decay
  mode, $h^\pm \to \mu^\pm \nu_\mu$, has evaded searches for lighter charged
  Higgses; see Ref.~\cite{Olive:2016xmw, Khachatryan:2015qxa,
    Aad:2014kga}. It is also possible that limits on supersymmetric scalar
  muons decaying as $\tilde\mu \to \mu + \etmiss$ require
  $M_{h^\pm} > 95\,\gev$~\cite{Olive:2016xmw}. A mass this large does not
  affect our results in Tables~1 and~2.}

The range of~$\alpha$ in Eq.~(\ref{eq:alphrange}) may seem unnaturally
small. But it is not our model's purpose to be devoid of all
fine-tuning. Certainly not with the enormous renormalizations of the Higgs
boson masses in this or any such model. The point here is that we can choose
the model's parameters to be consistent with ALEPH's
$Z \to \bar bb \mu^+\mu^-$ and Higgs-decay data, and so we will.

In option~2, we will see that it would be desirable to have
$M_{\eta_A} < M_Z - M_h$ there. But this is also excluded by the simultaneous
requirements of a stable Higgs potential, perturbative $|\lambda_i| < 4\pi$,
$0.8 < \cos(\beta - \alpha)$ --- a generous lower bound given the
$H \to WW^*$ signal strength, and achieving the branching
ratio~(\ref{eq:BZbbmm}) for the ALEPH signal.

In option~1, with $M_{\eta_A} \cong M_h \cong 30\,\gev$, the plausible origins
of the dimuon signal at ALEPH are $Z \to h\eta_A$ with $h \to \mu^+\mu^-$ and
$\eta_A \to \bar bb$, {\em and} vice-versa. There are also the
``Higgsstrahlung'' processes (1) $Z \to Z^* + h$ with $Z^* \to \bar bb$ and
$h \to \mu^+\mu^-$ and (2) $Z \to \bar bb$ with $b(\bar b) \to b(\bar b) + h$
or $\eta_A$ and $h/\eta_A \to \mu^+\mu^-$.
 
In option~2, with $\eta_A$ and $H$ nearly degenerate at $125\,\gev$, the only
kinematically plausible candidates for ALEPH are the Higgsstrahlung processes
with $h$-radiation. Their contribution to the $Z \to \bar bb \mu^+\mu^-$
branching ratio is tiny, $\simle 5\times 10^{-10}$ for any $v_2 < 35\,\gev$,
mainly because of suppression by the off-shell $Z$-propagator, the smallness
of $\sin(\beta - \alpha)$, and the weak coupling of~$h$ to~$\bar bb$. So,
option~2 cannot explain the 30~GeV dimuon excess.

The only source of the excess in option~1 is $Z \to h\eta_A$ because the
Higgsstrahlung processes are still negligible. In the narrow-width
approximation, the decay rate is
\bea\label{eq:Zheta}
&& \Gamma(Z \to h\eta_A \to \bar bb\mu^+\mu^-) = \Gamma(Z \to h\eta_A)\left[B(h
  \to\bar bb)B(\eta_A \to \mu^+\mu^-) + (h \leftrightarrow \eta_A) \right],
\nn\\
&&{\rm where\,\,\,}  \Gamma(Z \to h\eta_A) = \frac{2\alpha_{EM}\, p^3}{3
  M_Z^2\sin^2 2\theta_W} \cos^2(\beta-\alpha),
\eea
with $p$ the momentum of $h$ in the $Z$~rest frame. For
$M_h = M_{\eta_A} = 30\,\gev$, this gives
$B(Z \to h\eta_A) = 0.0141 \cos^2(\beta-\alpha)$.\footnote{The question of
  what to do about an additional $\sim 1.4\%$ in the $Z$ width is discussed
  briefly in Sec.~5.} Tables~1 and~2 list quantities of interest for a range
of $v_2$ and other inputs, including the choice
$M_{h^\pm} = 65\,\gev$.\footnote{We used $m_s = 0.057\,\gev$,
  $m_c = 0.71\,\gev$ and $m_b = 2.96\,\gev$ at $M_h = 30\,\gev$.} There is no
difficulty choosing parameters that produce a branching ratio in the
neighborhood of the value $B(Z \to \bar bb \mu^+\mu^-) = 2.77\times 10^{-4}$
deduced from the ALEPH data~\cite{Heister:2016stz}. Note that, because of the
relative smallness of~$\sin\alpha$, most of the dimuon signal comes from
$h \to \mu^+\mu^-$.


\begin{table}[!t]
  \begin{center}{
  \small
  \begin{tabular}{|c|c|c|c|c|c|}
    \hline
 $v_2\,(\gev)$ & $\beta$ & $\alpha$ & $\mu_3^2\,(\gev^2)$ & $\lambda_1$
    & $\lambda_2$  \\ 
  \hline\hline
    10 & 0.04066& $-0.348\times 10^{-2}$ & 36.6  & 0.1293 & 
 $0.833 \times 10^{-2}$ \\
  \hline
    12.5 & 0.05084& $-0.435\times 10^{-2}$ & 45.7  & 0.1294 & 
 $0.833 \times 10^{-2}$ \\
   \hline
    15 & 0.06101& $-0.522\times 10^{-2}$ & 54.8  & 0.1295 & 
 $0.833 \times 10^{-2}$ \\
  \hline
    20 & 0.08139& $-0.695\times 10^{-2}$ & 72.9  & 0.1299 & 
$0.832 \times 10^{-2}$ \\
  \hline
  \end{tabular}}
\caption{The parameters in option~1 of the two-Higgs doublet model vs.~the
  vev $v_2=\sqrt{2} \langle\phi_2\rangle_0$.  The other input parameters are
  $M_h = M_{\eta_A} = 30\,\gev$, $M_H = 125\,\gev$, $M_{h^\pm} =
  65\,\gev$.
  The quartic coupling combination $\lambda_3 + \lambda_4$ is held fixed at
  $-3.00\times 10^{-3}$ and $\lambda_4$ is chosen so that
  $M_{h^\pm} = 65\,\gev$. The results are insensitive to
  $|\lambda_3 + \lambda_4| \simle 5\times 10^{-3}$.
  \label{tab:params}}
 \end{center}
 \end{table}
\begin{table}[!ht]
  \begin{center}{
  \small
  \begin{tabular}{|c|c|c|c|c|c|}
    \hline
 $v_2\,(\gev)$ & $B(h\to\mu^+\mu^-)$ &$B(h\to\bar bb)$ &
 $B(\eta_A \to\mu^+\mu^-)$ & $B(\eta_A \to\bar bb)$ &
 $B(Z \to \bar bb \mu^+\mu^-)$ \\ 
   \hline\hline
 10 & 0.9999  & $0.444\times 10^{-4}$ & 0.9923 & $0.667\times 10^{-2}$ & 
$0.950\times 10^{-4}$ \\
\hline
 12.5 & 0.9998  & $1.085\times 10^{-4}$ & 0.9814 & $1.613\times 10^{-2}$ & 
$2.297\times 10^{-4}$ \\
\hline
 15 & 0.9997  & $2.249\times 10^{-4}$ & 0.9622 & $3.286\times 10^{-2}$ & 
$4.673\times 10^{-4}$ \\
\hline
 20 & 0.9991  & $7.105\times 10^{-4}$ & 0.8889 & $0.09652$ & 
$1.367\times 10^{-3}$ \\
\hline
  \end{tabular}}
\caption{The principal branching ratios of $h$, $\eta_A$ and of $Z \to
  \bar bb \mu^+\mu^-$ vs.~the vev $v_2 = \sqrt{2}\langle\phi_2\rangle_0$ in
  option~1 of the two-Higgs doublet model. The other input parameters are
  $M_h = M_{\eta_A} = 30\,\gev$, $M_H = 125\,\gev$, $M_{h^\pm} =
  65\,\gev$. The results are insensitive to $|\lambda_3 + \lambda_4| \simle
  5\times 10^{-3}$.
  \label{tab:rates}}
 \end{center}
 \end{table}
%


\section{Predictions}

This model makes a number of predictions, some obvious, some not so, that we
enumerate here.
\begin{itemize}

\item[1.)] The dimuon signal in ALEPH and in other detectors, at LEP
  or at the LHC, will be observed only in $Z$-decay and almost exclusively
  in association with $\bar bb$.

\item[2.)] Dimuons from the signal will have a common production
  vertex. Those outside the signal region are due to semileptonic $b$-decays
  and will not.

\item[3.)] Signal dimuons have an isotropic $\cos\theta^*$ distribution. Its
  flat shape is modified to a hump when there is a cut of
  $p_T > 5$--$10\,\gev$ on both muons. In that case, {\em all of the signal
    lies in $\theta^*_T < \theta^* < \pi-\theta^*_T$}, where $\theta^*_T$ is
  an increasing function of the $p_T$ cut. The effect is illustrated in
  Fig.~5 for LEP, where the $Z$ is produced at rest, and for the LHC, where
  the $Z$ tends to be produced with low~$p_T$ and a large boost. Note that
  the histograms are normalized to unit area so that, for a large $p_T$ cut,
  little signal data remains.
\begin{figure}[!ht]
 \begin{center}
\includegraphics[width=2.65in, height=2.60in]{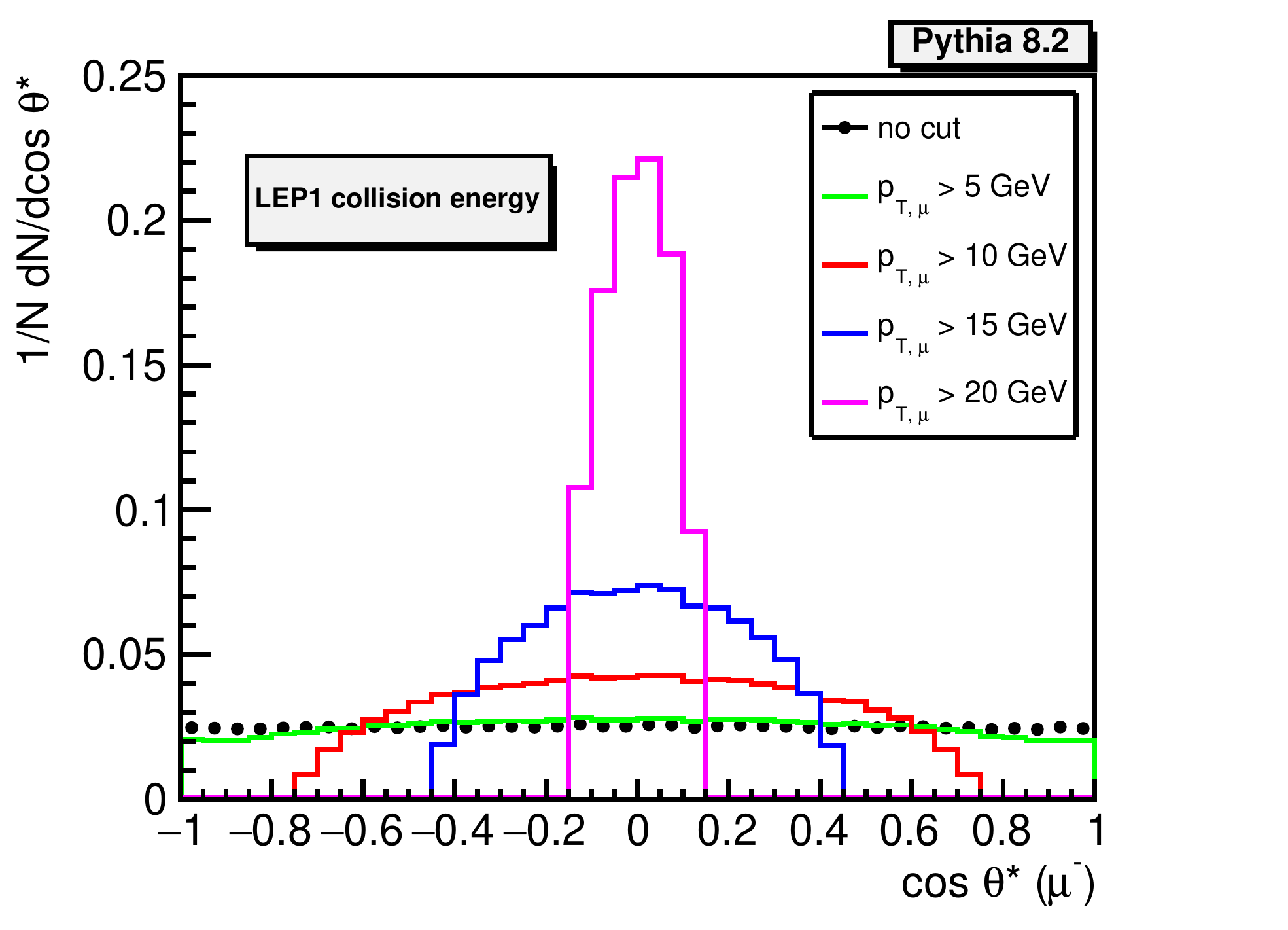}
\includegraphics[width=2.65in, height=2.60in]{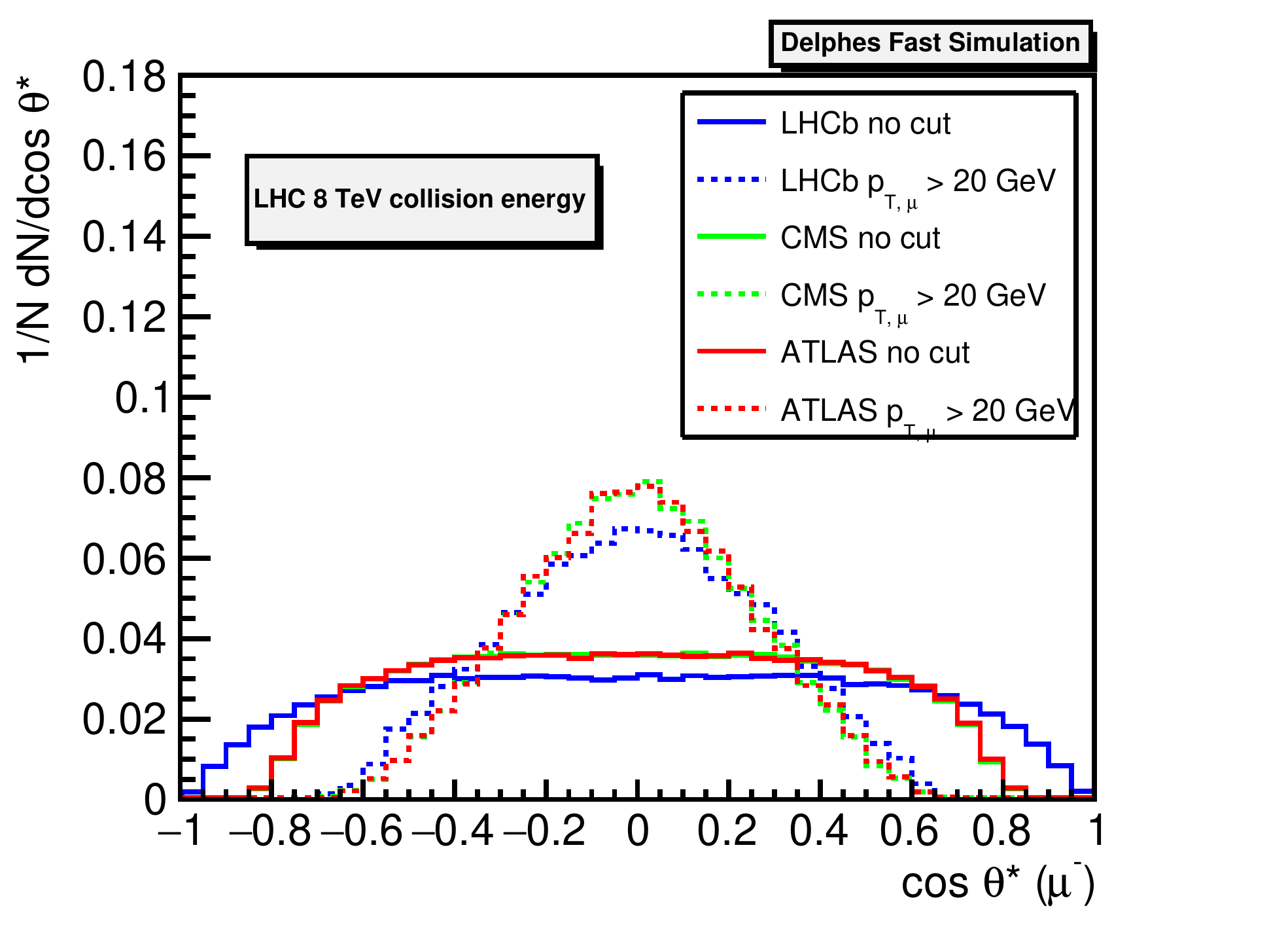}
\caption{The $\cos\theta^*$ distribution of the $\mu^-$ in
  $h,\eta_A \to \mu^+\mu^-$ as a function of the $p_T$ cut on each muon for
  LEP (left), where the $Z$ is produced at rest, and the LHC for collisions
  at $8\,\tev$ (right), assuming negligible $p_T(Z)$. From A.~Heister,
  private communication.}
  \label{fig:Figfab}
 \end{center}
 \end{figure}

\item[4.)] In our model, signal dimuons will {\em not} have a strong tendency
  to be close to the $b$-jets in the $Z$-boson's rest frame. We have checked
  that this obvious kinematical fact is true in any model in which
  $Z \to XY$ with $X \to \mu^+\mu^-$ and $Y \to \bar bb$, for $X,Y$ with
  spin-zero or one. This is in contradiction with the ALEPH data for which,
  when $M_{\mu^+\mu^-} \sim 30\,\gev$, the minimum angle between a muon and a
  leading jet is always less than $15^\circ$ degrees~\cite{Heister:2016stz}.
  We have no explanation for this difference. On the other hand, at the LHC,
  the rather large $Z$-boost makes the signal as well as semileptonic
  background muons less isolated. This tendency is stronger at $13\,\tev$
  than at $8\,\tev$. If muon isolation is an important signal criterion at
  $13\,\tev$, it may be possible to enhance it by selecting $Z+\jet$
  production. According to Ref.~\cite{Aaboud:2017hbk}, approximately 15\% of
  $Z$-production at $13\,\tev$ is accompanied by one jet with
  $p_T > 30\,\gev$.

\item[5.)] In the dimuon signal region, the $\bar bb$ invariant mass should
  have a significant excess near $M_{\eta_A}$, nominally $30\,\gev$ in our
  model.

\item[6.)] Charged Higgses, $h^\pm$, decay mainly to $\mu^\pm \nu_\mu$.
  Light charged Higgses may not have been excluded in this mode by previous
  searches~\cite{Olive:2016xmw,Khachatryan:2015qxa, Aad:2014kga}. If they
  were, they need to be heavier than $M_H/2 = 62.5\,\gev$. If they are
  excluded by LEP searches for supersymmetric scalar muons, they must be
  heavier than $95\,\gev$~\cite{Olive:2016xmw}. They are most readily sought
  in $\gamma^*,\, Z^* \to h^+ h^- \to \mu^+ \mu^- +\emiss$ in LEP-2 data and
  at the LHC and in
  $W^{*\,\pm} \to h/\eta_A + h^\pm \to \mu^+ \mu^- \mu^\pm \etmiss$ at the
  LHC.

\item[7.)] If $\alpha \ll \beta$, as in Table~1, $H(125)$ couples only weakly
  to $\mu^+\mu^-$, so this decay mode may be unobservably small.

\item[8.)] There will be no observable $30\,\gev$ excess in $M_{e^+e^-}$ in
  $Z \to \bar bb e^+ e^-$ events.

\end{itemize}

An interesting question is how to tell $h$ from $\eta_A$. The answer is not
obvious if they are nearly degenerate at $30\,\gev$. Another question for
which we have no ready answer is how to determine the mixing angles $\alpha$
and $\beta$ other than by naive fitting.

\section{A No-Go Theorem?}

To account for the apparently exclusive appearance of the 30~GeV dimuon
excess in association with $Z \to \bar bb$, we used a 2HDM in which the
Yukawa couplings of the second Higgs doublet $\phi_2$ involve only the muon
and electron doublets. The Yukawa couplings of the $\tau$ and quark doublets
are to $\phi_1$.\footnote{We remind the reader that this setup induces no
observable charged-lepton flavor violation.} Only option~1 with
$Z \to h\eta_A \to \bar bb \mu^+ \mu^-$ can explain the rate of the ALEPH
dimuon. In our model, for parameters that give $B(Z \to \bar bb \mu^+ \mu^-)$
in the vicinity of Eq.~(\ref{eq:BZbbmm}), we have
$B(x \to \mu^+\mu^-)/B(x \to \bar bb) \sim 10^2$ for $x = \eta_A$ and
$\sim 10^4$ for $x = h$. This makes
$B(Z \to h \eta_A \to 4\mu) \simeq 0.014$, 3300~times larger than its
measured value of $4.2 \times 10^{-6}$~\cite{Olive:2016xmw}.

We have considered several modifications of our model that decrease the
branching ratios of $h$~and~$\eta_A$ to $\mu^+\mu^-$ while increasing the
$\bar bb$ yield. We already mentioned that, for small~$\beta$ and $\alpha$,
$M_{\eta_A}$ cannot be much different from $M_h$. In our model,
$B(h \to \mu^+ \mu^-) B(\eta_A \to \bar bb) \propto \cos^2\alpha/\cos^2\beta$
and
$B(h \to \bar bb) B(\eta_A \to \mu^+ \mu^-) \propto \sin^2\alpha/\sin^2\beta$,
with $\sin\alpha/\sin\beta \sim 0.1$. So, the next simplest thing we
considered was to decrease $\cos\beta$. But since several production $\times$
decay-rate signal strengths of the Higgs boson~$H$ are also proportional to
$\cos^2\alpha/\cos^2\beta$, their measured values would no longer agree with
the SM expectation of unity. If we counter this by increasing~$\alpha$, with
$\beta-\alpha$ still small and $B(Z\to h\eta_A \to \bar bb \mu^+\mu^-)$ still
in the $10^{-4}$ range, we find again that the decay rates for $H \to hh$,
$\eta_A\eta_A$ are many 10's of MeV. Further, when increasing $\beta$, other
conflicts may arise, e.g., with $b \to s\gamma$ mediated by $h^\pm$-exchange.

In the context of a 2HDM, we also tried to ameliorate the 4-muon problem with
the Branco-Grimus-Lavoura (BGL) mechanism~\cite{Branco:1996bq,
  Botella:2014ska, Bhattacharyya:2014nja} to dilute
$B(h,\eta_A \to \mu^+\mu^-)$. The BGL scheme admits Higgs-induced
flavor-changing neutral current interactions (FCNC) through a softly-broken
$U(1)_\phi$ symmetry that allows a set of quarks with the same electric
charge and color to couple to and get mass from both Higgs
doublets~\cite{Glashow:1976nt}. The resulting FCNC involve only the quark
masses and elements of the CKM matrix~$V$. If the third generation is treated
differently than the first two, the FCNC are suppressed by factors of
$V_{3i}$ or $V_{i3}$ and they can be sufficiently small even for Higgs masses
much less than the 100-1000 TeV scale ordinarily required by $|\Delta S| = 2$
and $|\Delta B| = 2$ constraints.

We considered plausible alternatives in which the Yukawa couplings to
$\phi_1$ and $\phi_2$ of one type of quark, up or down, have the form (here
$\times$ denotes a nonzero entry):
\be\label{eq:GamBGLud}
\Gamma_1 = \left(\begin{array}{ccc} 0 &0 &0\\ 
                                         0 &0 &0\\ 
                                         \times &\times
                                          &\times\end{array}\right),\quad
\Gamma_2  = \left(\begin{array}{ccc} \times &\times &\times\\ 
                                         \times &\times &\times\\ 
                                         0 &0 &0\end{array}\right),\quad {\rm
                                       or\,\,vice-versa,}
\ee
while those of the other type are
\be\label{eq:GamBGLdu}
\Gamma_1 = \left(\begin{array}{ccc}  0 &0 &0\\ 
                                         0 &0 &0\\ 
                                         0 &0 &\times\end{array}\right),\quad
\Gamma_2  = \left(\begin{array}{ccc} \times &\times &0\\ 
                                         \times &\times &0\\ 
                                         0 &0 &0\end{array}\right),\quad {\rm
                                       or\,\,vice-versa}.
\ee
The Yukawa textures of the leptons are the same as those displayed in
Eq.~(\ref{eq:GamBGLdu}). The ``vice-versa'' textures in these two equations
are excluded for the $u$~and the $d$-sectors. If used for the $u$-sector,
they imply $\bar tt$ couplings to $h,\eta_A$ of $\CO((m_t/v)\cot\beta)$ and
to $H$ of $\CO((m_t/v)\tan\beta)$. This ruins the agreement of the $H$-signal
strengths with the SM and implies that by far the dominant decay modes of
$h,\eta_A$ are to two gluons! Using them for the $d$-sector implies, among
other things, that
$B(h,\eta_A \to \bar bb)/B(h,\eta_A \to \mu^+ \mu^-) = (3m_b/m_\mu)^2$, so
that it is impossible to have
$B(Z \to h\eta_A \to \bar bb \mu^+ \mu^-) \sim 10^{-4}$.

For the displayed textures, the $|\Delta\, {\rm Flavor}| = 2$ interactions
induced by light $h$ and $\eta_A$ exchange are all very small because of a
near cancellation between the two terms as well as the suppression by
$V_{3i}$ or $V_{i3}$.\footnote{This cancellation $h$-$\eta_A$ was noted in
  Ref~\cite{Botella:2014ska} but there was no reason for $M_h = M_{\eta_A}$
  in that paper.} However, the textures in Eq.~(\ref{eq:GamBGLud}) for the
$d$-sector give $B(B_{d,s} \to \mu^+\mu^-)$ which are $10^7$ times larger
than their experimental upper limits. Furthermore, for $\Gamma_1^d$ and
$\Gamma_2^d$ as displayed in either of these two sets of textures,
$B(h,\eta_A \to \bar ss)$ is almost as large as for $\mu^+\mu^-$. Such a
large rate would have been captured in the ALEPH data for which the $b$-tag
was inverted (see Fig.~4).

For $u$-sector FCNC with Eq.~(\ref{eq:GamBGLud}),
 \be\label{eq:Dmm}
 B_{BGL}(D^0 \to \mu^+\mu^-) = \left[\frac{f_D M_{D^0}^2 m_\mu V_{ub}
     V^*_{cb}}{v^2 M_h^2 \sin^2\beta}\right]^2 \frac{M_{D^0}}{16\pi
   \Gamma_{D^0}} = 1.65\times 10^{-10},
 \ee
 for $\sin\beta = 0.05$ and $f_D = 212\,\mev$. This is to be compared to the
 limit $B(D^0 \to \mu^+ \mu^-) < 6.2\times 10^{-9}$~\cite{Olive:2016xmw}.
 However, $B(h,\eta_A \to \bar cc) = 0.993$,
 $B(h,\eta_A \to \mu^+\mu^-)= 0.733\times 10^{-2}$ and
 $B(h,\eta_A \to \bar bb) = 1.08\times 10^{-4}$, and that kills off this BGL
 version of our model. To sum up, then, none of the BGL models alleviates the
 $Z \to 4\mu$ problem without introducing others that are just as bad and,
 generally, even give up on explaining the $Z \to \bar bb + 30$-GeV dimuon rate
 in ALEPH.

 Another way to look at the $Z \to 4\mu$ problem is that the ALEPH signal
 accounts for only about 2\% of the $Z \to h\eta_A$ decay rate predicted by
 the model for the nominal case $M_h \simeq M_{\eta_A} \simeq 30\,\gev$.
 Where is the other 98\% going if not into four muons? Can it be into quarks?
 Our foray into BGL models suggests not but, at bottom, this is an
 experimental question of determining individual $Z$-decay branching
 ratios. Can it be going into something invisible?  How can we account for
 $B(Z \to h\eta_A) \simeq 1.4\%$ when the $Z$ width is measured to 0.1\% and
 the $Z \to$ invisibles width implies the number of neutrinos is
 $2.92\pm 0.05$~\cite{Olive:2016xmw}?

These sorts of problems would seem to infect any scalar-based model of the
ALEPH dimuon excess, especially because a novel scalar coupling to leptons
and quarks must involve Higgs multiplets beyond the Standard Model.
We have not considered vector-based models in much depth. Our prejudice is
that all vector bosons are gauge bosons. One example would be to set
$\mu_3^2 = 0$, gauge $U(1)_\phi$, and absorb $\eta_A$ in the corresponding
gauge boson. However, assuming no other fermions than those in the SM, it is
straightforward to see that canceling all gauge anomalies is possible only if
the $U(1)_\phi$ hypercharge of each fermion is proportional to its
weak-$U(1)$ hypercharge. So, a more complicated setup is needed. What
guidance does the data give us that might evade the $4\mu$ problem?

Are we faced with a no-go theorem for explaining the ALEPH dimuon excess? It
has been said that there are no no-go theorems.\footnote{J.D.~Bjorken,
  private communication to K.L., 1977.}  Evading one is ``simply'' a matter
of changing the assumptions of the theorem. But what assumptions should we
change? If the narrow 30~GeV ALEPH dimuon, its apparent $Z$-boson source, and
its association with $\bar bb$ production are confirmed in data from other
LEP or LHC experiments, the challenge will be to theorists to account for it
within the constraints of the PDG book~\cite{Olive:2016xmw}.

\section{Acknowledgments}

We are grateful for valuable conversations with and comments from Kevin
Black, Tulika Bose, Estia Eichten, Shelly Glashow, Howard Georgi, Arno
Heister, Adam Martin, Chris Rogan, Martin Schmaltz, Jesse Thaler and Bing
Zhou.


\bibliography{Aleph-Dimuon}
\bibliographystyle{utcaps}
\end{document}